\documentclass[12pt]{iopart}
\usepackage{graphicx}
\begin{document}

\title[Thermally controlled dissipative soliton crystals in optical microresonators]{Soliton crystals in optical Kerr microresonators in the presence of thermo-optic effects}

\author{B. Azah Bei Cho$^1$, I. Ndifon Ngek$^1$ \& Alain M. Dikand\'e$^{1}$\footnote{Corresponding author}}

\address{$^1$ Laboratory of Research on Advanced Materials and Nonlinear Sciences (LaRAMaNS), Department of Physics, Faculty of Science, University of Buea PO Box 63, Buea, Cameroon}

\ead{dikande.alain@ubuea.cm}
\vspace{10pt}
\begin{indented}
\item[]
\end{indented}

\begin{abstract}
The influence of thermo-optic effects on shape profiles of soliton crystals in optical Kerr microresonators is investigated. The study rests on a model that consists of the Lugiato-Lefever equation, coupled to the one-dimensional heat diffusion equation with a source term proportional to the average power of the optical field. Using appropriate variable changes the model equations are transformed into a set of coupled first-order nonlinear ordinary differential equations. These equations are solved numerically with emphasis on the influence of thermo-optic effects on the amplitude and instantaneous frequency of the optical field, as well as on the temperature profile in the microresonator cavity. It is found that thermo-optic effects do not prevent soliton crytals from forming in optical Kerr microresonators, however, a strong thermal detuning will decrease the soliton-crystal amplitude. The model predicts a temperature profile in the microresonator cavity which is insensitive to the specific spatio-temporal profile of the soliton crystal propagating in the microresonator, a feature peculiar to the model.  
\end{abstract}

%
%
%
%
%

\section{Introduction}
Optical Kerr microresonators have attracted a great deal of interest in the recent past \cite{a1,a2,a3,a4,a5,a6,a61,a62}, due to their ability to generate wavetrain patterns that can cover a spectral region over an octave \cite{a7,a8} while operating in a low--noise and phase--stable configuration. These wavetrain patterns are particular in that they exhibit features associated with their periodic structures, resulting from the crystallization of a large number of identical optical waves multiplexed to form a comb-like wave pattern with a well defined periodicity \cite{a9,a10,a11,a12,a13}. Soliton crystals \cite{d1,d2} are a specific class of periodic wavetrains observed in optical microresonators, where they are favoured by the competition between a Kerr nonlinearity and the spectral dispersion prevailing in the optical microresonator. This competition triggers the generation of frequency combs when the free spectral range of the microresonator grows with frequency due to the anomalous dispersion, inducing multiple parametric resonant four-wave mixings that will ultimately lead to a pattern of evenly spaced frequency lines \cite{fw1}. When the complete spectrum of microresonator's resonant frequencies contributes to the parametric four-wave mixing processes, the nonlinear dynamics of the optical field gives rise to a comb with a single free spectral range frequency spacing identified as soliton crystals. Soliton crystals (i.e. soliton-comb structures) find widespread applications in optical metrology such as high-resolution spectroscopy, atomic clocks, optical sensings including light detection and ranging (LIDAR) \cite{lid}, exoplanet exploration \cite{lid1}, optical frequency synthesis \cite{lid2} and so on. \par
In practical contexts, soliton combs are produced by feeding Kerr microresonators with continuous-wave (CW) lasers which promote spectral side-bands through four-wave mixing cascades \cite{fw1}. Although experiments suggest that soliton-comb structures are more likely to form on the red-detuned side of microresonator’s pump resonance mode \cite{10,11,12}, it has been observed that in this spectral region thermo-optical effects can be so sizable that the soliton crystal stability is hampered \cite{10,11,12}. To understand the influence of thermo-optic effects on the dynamics of soliton-comb structures in Kerr optical microresonators, recently a theoretical model was proposed \cite{mod1,mod2}. The model consists of the Lugiato-Lefever equation for the optical field propagation, with a term accounting for thermal detuning associated with the temperature variation in the microresonator cavity. The temperature variation in this model is governed by the one-dimensional heat equation, in which the heat source term is proportional to the average power of the optical field within the microresonator cavity. In ref. \cite{mod1} the authors examined the influence of thermo-optic effects on the stability of single-soliton and soliton-crystal structures in the ring-shaped optical Kerr microresonator, thus they established that in the absence of thermo-optic effects the stable regions for single solitons and soliton crystals partially overlap. However, when thermal effects are included via the thermal detuning term in the Lugiato-Lefever equation, the stability region of single solitons separates completely from the one of soliton crystal. They inferred that the differences in stability features were an evidence of the effectiveness of backwards-detuning in obtaining single-soliton structures in optical Kerr microresonators. But to gain a more detailed picture of the influence of thermo-optic effects on soliton-comb structures in the ring microresonator, it is more interesting to examine profiles of the soliton combs by solving the Lugiato-Lefever equation and the heat equation coupled to it \cite{mod1,mod2}. \par 
In the present study we wish to investigate solutions to the model proposed in ref. \cite{mod1,mod2}, focusing our attention on the nonlinear regime where soliton structures are expected to be dominant. To proceed, we adopt a specific ansatz \cite{fand} for the optical field, which describes an electromagnetic field with real amplitude and real phase moving on a stationary frame at a constant propagation speed. With the help of this specific ansatz, the system dynamics is mapped onto a set of coupled first-order ordinary differential equations, which shall be solved numerically for some combinations of characteristic parameters of the model. In particular we analyze the influence of variations of the thermo-optic coefficients and the linear loss, on the amplitude and instantaneous frequency of the optical field propagating in the Kerr microresonator.

\section{The model and singular solutions}
We are interested in the nonlinear dynamics of an electromagnetic field propagating within the cavity of a ring-shaped optical Kerr microresonator, in the presence of thermo-optic effects. Thermo-optic effects in this physical system stem from heat deposition in the microresonator cavity during the optical field roundtrips, which generates an extra detuning process associated with heat and referred to as thermal detuning, represented by a temperature-dependent angular frequency $\Phi(T)$ of the microresonator mode when the cavity temperature is $T$ \cite{mod1,mod2}. For this physical system, the Lugiato-Lefever equation describing the propagation of the electromagnetic field in the optical Kerr microresonator is expressed \cite{mod1,mod2}:
\begin{equation}
\frac{\partial \psi}{\partial t}= \frac{i\beta}{2}\,\frac{\partial^2\psi}{\partial x^2} +i\gamma\vert\psi\vert^2\psi -\Big[ \lambda +i\Big(\alpha +\Phi(T)\Big)\Big]\psi + F, \label{eq1}
\end{equation}
where $\psi$ is the envelope of the slowly-varying electromagnetic field, $t$ is time and $x$ is the roundtrip variable. $\beta$ is the group-velocity dispersion coefficient, $\gamma$ is the nonlinear coefficient, $\lambda$ is the linear cavity loss, $\alpha$ is the normalized angular frequency detuning between the cavity resonance and the pump laser, $\Phi(T)$ is the thermal detuning and $F$ is the pump field. \par The thermal detuning $\Phi(T)$ is a function of temperature and is assumed to change with time, due to thermo-optic effects. Following a previous study \cite{mod1} we write $\Phi(T)=q\,\Delta T$, where $q$ is real and proportional to the thermo-optic coefficient $\partial n/\partial T$. $\Delta T=T-T_0$ is the difference between the cavity temperature $T$ and the room temperature $T_0$. The change with time of the cavity temperature is governed by the one-dimensional heat equation, such that the time evolution of the thermal detuning $\Phi(T)$ is governed by the following two-level type equation \cite{mod1,mod2}:
\begin{equation}
\frac{\partial \Phi}{\partial t}=A P-B\Phi. \label{eq2}
\end{equation}
Here $P=L^{-1}\int_{CL}{dx\,\vert \psi\vert^2}$ is the average power of the optical field in the microresonator cavity, where the integral is taken over the cavity length (CL, or circumference in the specific context of microresonator with circular cavity) corresponding to the roundtrip period $L$. The physical meanings and typical values of the coefficients $A$ and $B$ for some concrete physical contexts, are detailed in refs. \cite{mod1,mod2}. Therefore, here we shall simply refer to them as thermal coefficients. In ref. \cite{mod1} a linear stability analysis of the above model was carried out, from this previous study it emerged that the main signature of thermo-optic effects was a change in the effective detuning. Typically the change in the effective detuning was shown to be smaller for single solitons than for soliton crystals. It was also established that in the absence of thermo-optic effects the stability regions for single-soliton and soliton-crystal structures are almost the same, and when thermal effects are taken into consideration the stability region of single solitons separates from the stability region for soliton crystals. The difference in stability features was coined as a signature of thermo-optic effects in favouring single-soliton structures over soliton-crystal structures in specific physical contexts \cite{mod1}. \par
We seek to examine the influence of thermo-optic effects on the profile of soliton structures propagating in the microresonator. In this purpose let us consider an optical field $\psi(z,t)$, describing an electromagnetic field with a real amplitude and real phase both varying in a frame of reference $\tau=x-\vartheta t$, where $\vartheta$ is the propagation speed \cite{fand}. An ansatz corresponding to such envelope is:
\begin{equation}
\psi(x,t)=a(\tau)\,\exp\Big\lbrace i\big[\phi(\tau)-\omega t\big]\Big\rbrace, \label{eq3}
\end{equation}
where $a(\tau)$ and $\phi(\tau)$ are the amplitude and phase of the optical field respectively, while $\omega$ is a characteristic frequency. Substituting eq. (\ref{eq3}) in eq. (\ref{eq1}) with the external pump $F$ expressed as $F=f_0\,e^{-i\phi_0}$, where $f_0$ and $\phi_0$ are both real, after separation of real from imaginary parts we obtain:
\begin{eqnarray}
\Big(\lambda &+& \frac{\beta}{2}M_{\tau}\Big)a + \Big(\beta M-\vartheta\Big)y -f_0\cos\Big(\phi+\phi_0-\omega t\Big)=0, \label{eq4a} \\
\Big(\omega &+& \vartheta M-\alpha -\Phi -\frac{\beta}{2}M^2\Big)a +\frac{\beta}{2}y_{\tau} +\gamma a^3 -f_0\sin\Big(\phi+\phi_0-\omega t\Big)=0, \nonumber\\ \label{eq4b} \\
y&=&a_{\tau}, \qquad M=\phi_{\tau}.
\end{eqnarray}
 We can rewrite the above set together with the heat equation (\ref{eq2}), as a system of five coupled first-order ordinary differential equations i.e.:
\begin{eqnarray}
a_{\tau}&=&y, \label{eq5aa}\\ \phi_{\tau}&=&M, \label{eq5ab}\\
y_{\tau}&=& M^2a+\frac{2}{\beta}\Big[\Big(\alpha + \Phi-\omega -\vartheta M\Big)a -\gamma a^3\Big] + \frac{2}{\beta}\,f_0\sin\Big(\phi+\phi_0-\omega t\Big), \label{eq5a} \\
M_{\tau}&=& -\frac{2\lambda}{\beta} +\frac{2}{\beta}\Big(\vartheta-\beta M\Big)\frac{y}{a} + \frac{2}{\beta}\,\frac{f_0}{a}\cos\Big(\phi+\phi_0-\omega t\Big), \label{eq5b} \\
\Phi_{\tau}&=& -\frac{A}{\vartheta}P + \frac{B}{\vartheta}\Phi. \label{eq5c} 
\end{eqnarray}
Equations (\ref{eq5aa}) to (\ref{eq5c}) form a system of five coupled first-order nonlinear ordinary differential equations, whose solutions provide the essential information on the dynamics of the model introduced in refs. \cite{mod1,mod2}. Let us first start with the fixed points of the dynamical system (\ref{eq5aa})--(\ref{eq5c}), which are singular solutions corresponding to a regime of steady motion for the optical field. Such a regime is characterised by $y=0$, $y_{\tau}=0$, $\vartheta=0$ and the instantaneous frequency of the optical field, $M=0$; and we  find that
\begin{eqnarray}
\frac{f_0}{a}\cos\Big(\phi+\phi_0 - \omega t\Big)&=&\lambda, \label{ap1} \\
\frac{f_0}{a}\sin\Big(\phi+\phi_0 - \omega t\Big)&=&\gamma a^2 + (\omega -\Lambda). \label{ap2}
\end{eqnarray}
We can square the two equations  and then sum them, leaving us with the polynomial of power six in $a$ given by 
\begin{eqnarray}
\omega^2&+&2\Big(\gamma a^2-\Lambda\Big)\omega +\lambda^2 +\Big(\Lambda - \gamma a^2\Big)^2 - \Bigg(\frac{f_0}{a}\Bigg)^2=0, \label{eq6a} \\
\Lambda&=&\alpha + \Phi_e, \label{eq6}
\end{eqnarray}
where $\Phi_e$ is the thermal detuning at equilibrium. Note that $\Phi_e$ is extracted from eq. (\ref{eq2}) at fixed point and reads:
\begin{equation}
\Phi_e=\frac{AP}{B},
\end{equation}
suggesting that at equilibrium, the cavity is at a temperature:
\begin{equation}
T_e=T_0 -\frac{AP}{Bq}. \label{eq8}
\end{equation}
It is remarkable that since the temperature $T_e$ must be positive, one of the two thermal coefficients $A$ and $B$ should be negative, assuming that $q$ can only take on positive values. This implies that as we increase $q$, thermal effects should be enhanced and from formula (\ref{eq8}) we infer a relatively small difference between the room temperature and the cavity equilibrium temperature. In fact, this is not the scenario predicted in ref. \cite{mod1}, where it was considered that in addition to being negative, the coefficient $A$ must be proportional to $q$. It turns out that the equilibrium temperature $T_e$ of the microresonator cavity will instead depend mainly on the average power $P$ of the optical field, as well as on $B$ which should be typically always smaller than the absolute value of $A$ \cite{mod1}. \par
\begin{figure}\centering
\begin{minipage}[c]{0.32\textwidth}
\includegraphics[width=1.8in, height= 1.5in]{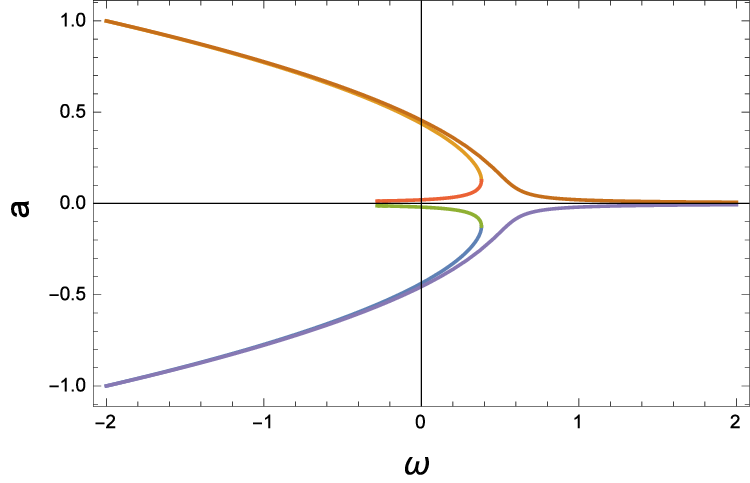}
\end{minipage}%
\begin{minipage}[c]{0.32\textwidth}
\includegraphics[width=1.8in, height= 1.5in]{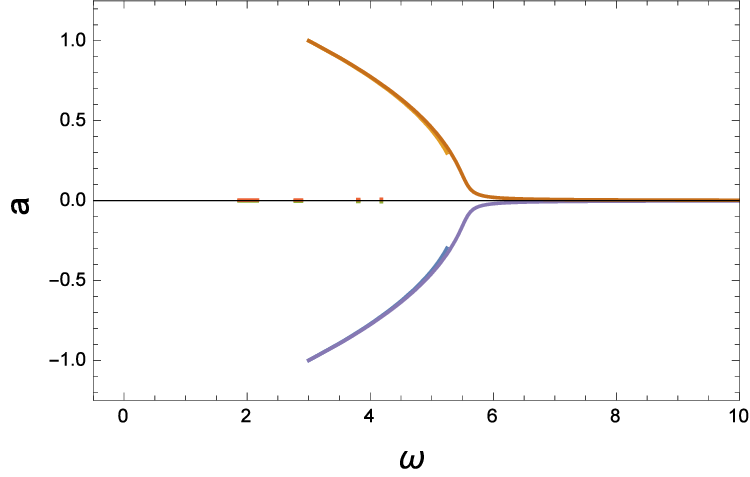}
\end{minipage}%
\begin{minipage}[c]{0.32\textwidth}
\includegraphics[width=1.8in, height= 1.5in]{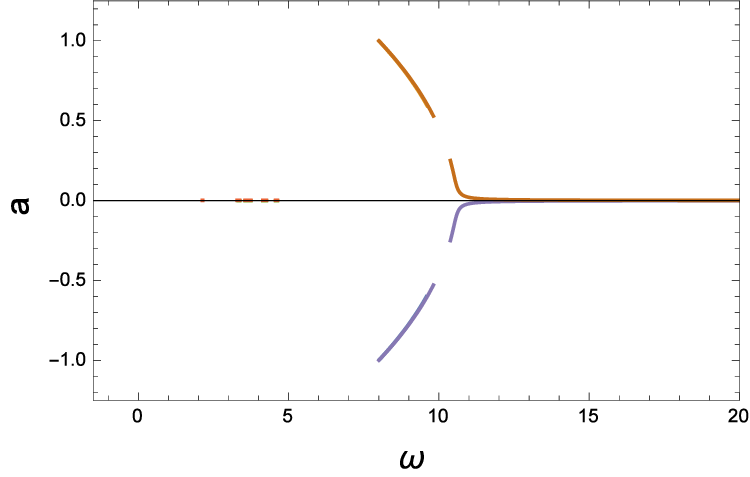}
\end{minipage}
      \caption{(Colour online) Fixed point solution for the field amplitude $a$, plotted as a function of the characteristic frequency $\omega$ for $\gamma=2.5$, $\lambda=-0.02$ and $f_0=-0.01$. From left to right graphs, values of $\Lambda$ are $0.5$, $5.5$ and $10.5$.}{\label{fig1a}}
  \end{figure}

\begin{figure}\centering
\begin{minipage}[c]{0.32\textwidth}
\includegraphics[width=1.8in, height= 1.5in]{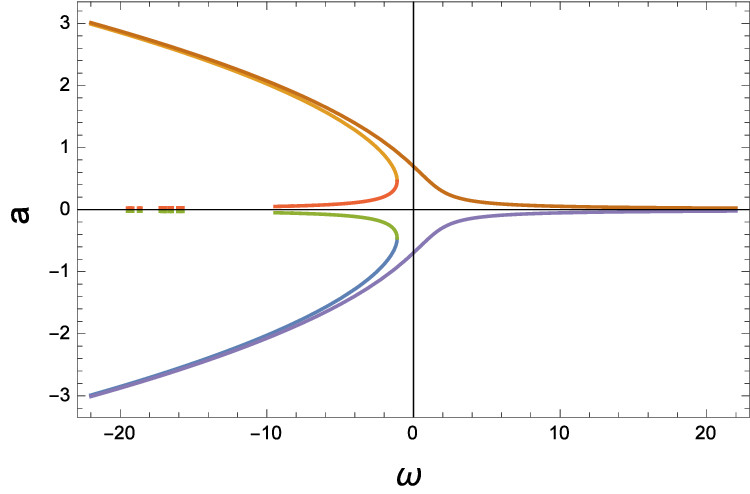}
\end{minipage}%
\begin{minipage}[c]{0.32\textwidth}
\includegraphics[width=1.8in, height= 1.5in]{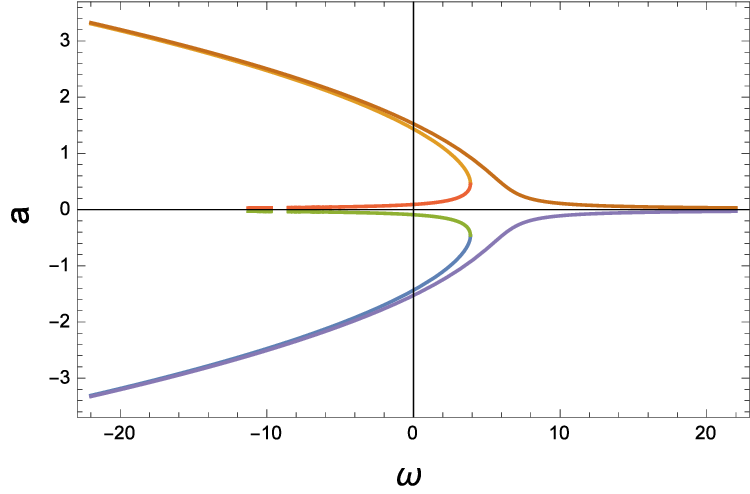}
\end{minipage}%
\begin{minipage}[c]{0.32\textwidth}
\includegraphics[width=1.8in, height= 1.5in]{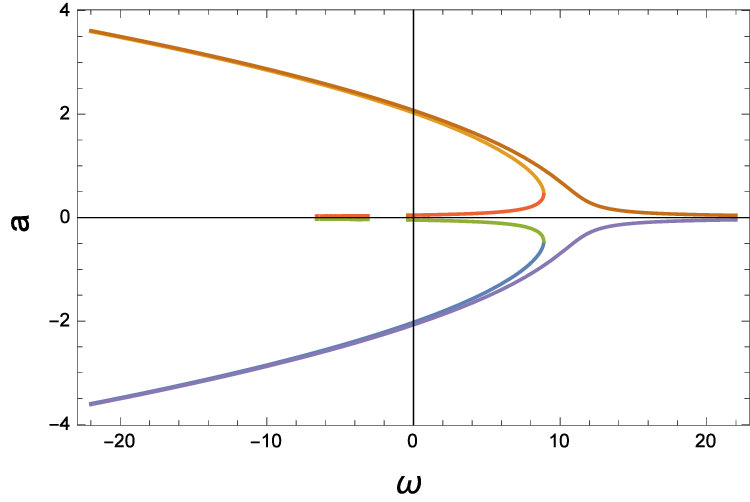}
\end{minipage}
      \caption{(Colour online) Fixed point solution for the field amplitude $a$, plotted a a function of the characteristic frequency $\omega$ for $\gamma=2.5$, $\lambda=-0.02$ and $f_0=-0.5$. From left to right graphs, values of $\Lambda$ are $0.5$, $5.5$ and $10.5$.}{\label{fig1b}}
  \end{figure}

 
The four possible real roots of the transcendental polynomial (\ref{eq6a}), corresponding to fixed points of the optical field amplitude $a$, are plotted in figs. \ref{fig1a} and \ref{fig1b} against $\omega$ for two different values of the optical pump amplitude $f_0$. In each figure we considered three values of the effective detuning $\Lambda$, which according to eq. (\ref{eq6}) are equivalent to changing the total detuning via change of the equilibrium thermal detuning $\Phi_e$.
  
 The left graph of fig. \ref{fig1a} corresponds to the smaller value of $f_0$ and a small value of the effective detuning $\Lambda_e$. The graph indicates that the region of stability of the optical field is confined in the range of relatively small $\omega$. As $\omega$ is increased, the fixed-point value of the optical field amplitude decreases and vanishes asymptotically at large values of $\omega$. The two other graphs (i.e. the middle and right graphs), however, exhibit a distinct behaviour: we clearly see that the stability region of the optical field is extended to larger values of $\omega$, when the effective detuning $\Lambda$ is increased. Graphs of fig. \ref{fig1b} are qualitatively similar to those of fig. \ref{fig1a}, although the former graphs are  richer in points representing stable solutions in the $a-\omega$ plane than the latter ones (i.e. graphs of fig. \ref{fig1a}).
\section{Soliton solutions}
Let us turn to profiles of the optical field amplitude as it propagates within the microresonator cavity, under the influence of thermo-optic effects. In this respect we solve the set of coupled first-order nonlinear ordinary differential equations (\ref{eq5aa})-(\ref{eq5c}) numerically, using a fourth-order Runge-Kutta scheme. To simplify the analysis and without loss of generality, we shall express the heat equation in terms of the variable $r=(T-T_0)/T_0$. With this new variable, the set of coupled first-order nonlinear ordinary differential equations to be solved numerically reads:
\begin{eqnarray}
a_{\tau}&=& y, \label{eq9a1} \\
\phi_{\tau}&=& M, \label{eq9a2} \\
y_{\tau}&=& M^2a+\frac{2}{\beta}\Big[\Big(\alpha + qT_0\,r-\omega -\vartheta M\Big)a -\gamma a^3\Big] \nonumber \\ &+& \frac{2 f_0}{\beta}\sin\Big(\phi+\phi_0-\omega t\Big), \label{eq9a} \\
M_{\tau}&=& -\frac{2\lambda}{\beta} +\frac{2}{\beta}\Big(\vartheta-\beta M\Big)\frac{y}{a} + \frac{2f_0}{a \beta}\cos\Big(\phi+\phi_0-\omega t\Big), \label{eq9b} \\
r_{\tau}&=& -\frac{A}{qT_0\vartheta}P + \frac{B}{\vartheta}r. \label{eq9c} 
\end{eqnarray}
Obviously it is difficult, within the framework of the present study, to bring out the influence of the many characteristic parameters present in the set of coupled equations (i.e. $\alpha$, $\beta$, $q$, $T_0$, $\omega$, etc.), on numerical solutions to these equations. Therefore we stick to our primary objective and focus mainly on the influence of thermo-optic effects on the amplitude $a(\tau)$ and instantaneous frequency $M(\tau)$ of the optical field, two quantities that can enable us gain sufficiently rich insight into the problem. Concretely, we shall maintain the thermal coefficient $B$ fixed and vary the other thermal coefficient i.e. $A$. For each of the selected values of $A$ the linear cavity loss coefficient $\lambda$ will be varied, starting from zero.
\par
\begin{figure}\centering
\begin{minipage}[c]{0.32\textwidth}
\includegraphics[width=1.8in, height= 1.5in]{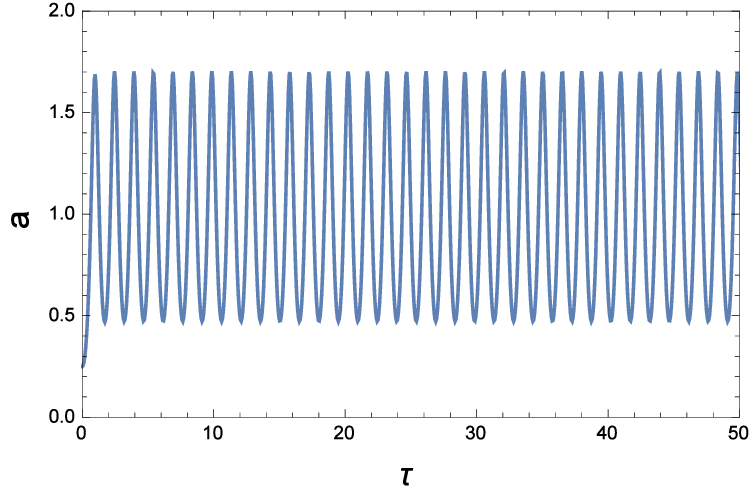}
\end{minipage}%
\begin{minipage}[c]{0.32\textwidth}
\includegraphics[width=1.8in, height= 1.5in]{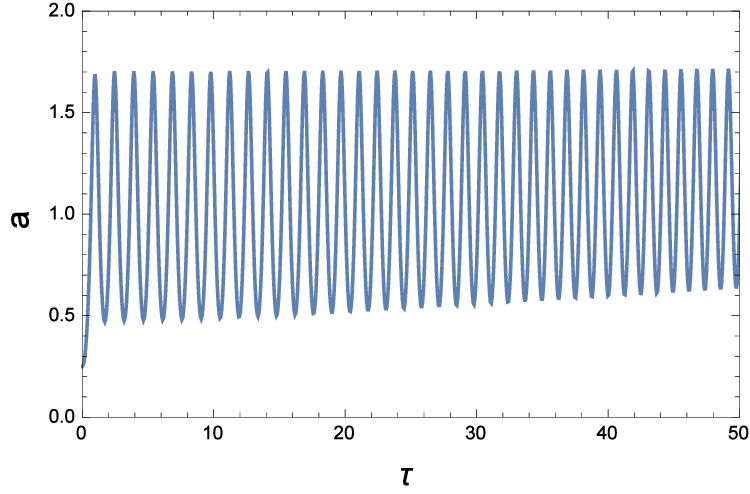}
\end{minipage}%
\begin{minipage}[c]{0.32\textwidth}
\includegraphics[width=1.8in, height= 1.5in]{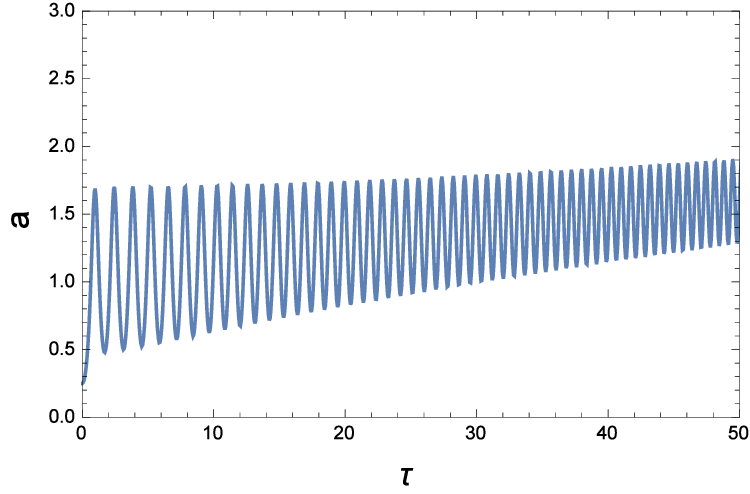}
\end{minipage}
\caption{(Colour online) The optical-field amplitude $a$ as a function of $\tau$ for $\gamma = 5.5$, $\beta=1.2$, $\alpha=0.5$, $\omega=5$, $\vartheta= - 0.08$, $P= 1.5$, $\phi_0 = -\pi$, $q=1$, $f_0=0.05$, $B = 0.2$, $A = 0.5$. Left graph is for $\lambda=0$, middle graph for $\lambda=-0.01$, and right graph for $\lambda=-0.05$.}{\label{fig2a}}
  \end{figure}

\begin{figure}\centering
\begin{minipage}[c]{0.32\textwidth}
\includegraphics[width=1.8in, height= 1.5in]{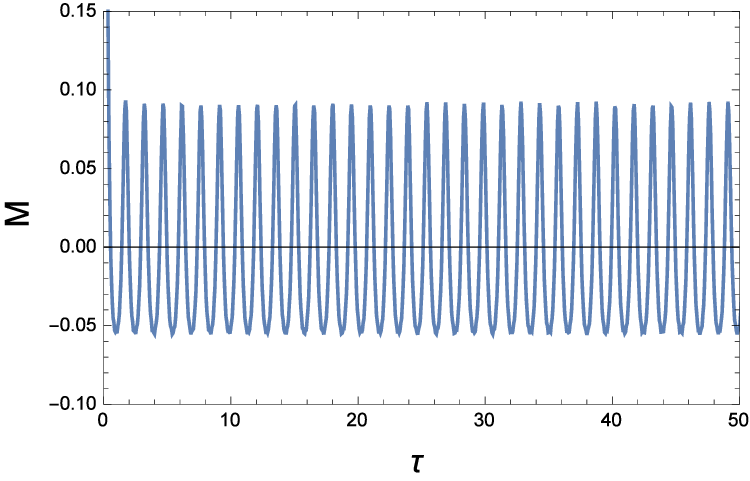}
\end{minipage}%
\begin{minipage}[c]{0.32\textwidth}
\includegraphics[width=1.8in, height= 1.5in]{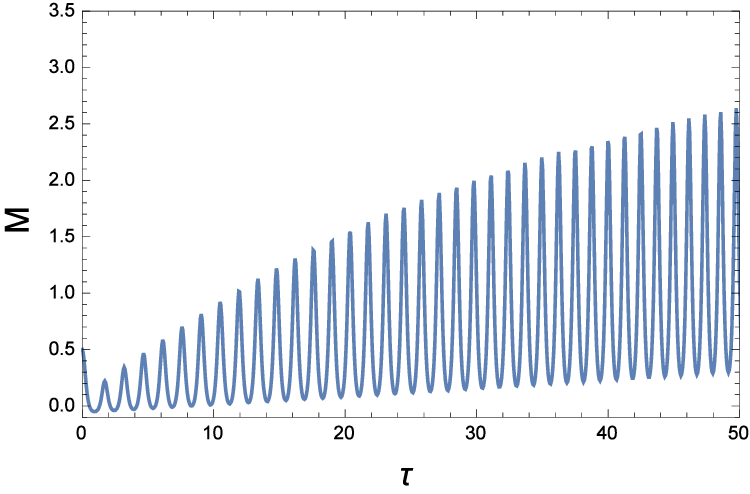}
\end{minipage}%
\begin{minipage}[c]{0.32\textwidth}
\includegraphics[width=1.8in, height= 1.5in]{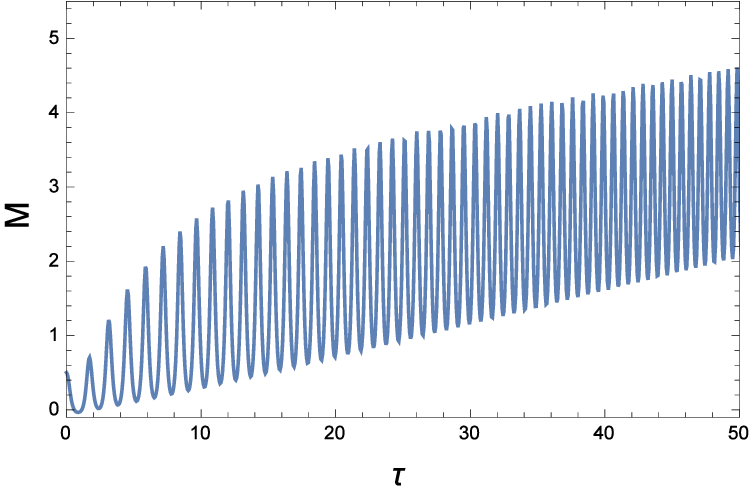}
\end{minipage}
\caption{(Colour online) Instantaneous frequency $M$ of the optical field, plotted a function of $\tau$ for $\gamma = 5.5$, $\beta=1.2$, $\alpha=0.5$, $\omega=5$, $\vartheta= - 0.08$, $P= 1.5$, $\phi_0 = -\pi$, $q=1$, $f_0=0.05$, $B = 0.2$,
$A = 0.5$. Left graph is for $\lambda=0$, middle graph for $\lambda=-0.01$, and right graph for $\lambda=-0.05$.}{\label{fig2b}}
  \end{figure}

 The three graphs in fig. \ref{fig2a} and fig. \ref{fig2b} represent the optical-field amplitude $a(\tau)$ and instantaneous frequency $M(\tau)$ respectively, against $\tau$ for $B=0.2$ and $A=0.5$. In each of the two figures the left graph corresponds to $\lambda=0$, the middle graph to $\lambda=-0.01$ and the right graph to $\lambda=-0.05$. In fig. \ref{fig2a} we observe that at zero cavity loss, the optical field deploys to form a regular periodic train of equal-amplitude solitons characteristic of a soliton-crystal pattern. As $\lambda$ becomes nonzero, the optical-field amplitude evolves being marked by an exponential damping, a signature of soliton crystals with dissipative amplitudes which we readily refer to as dissipative soliton crystals.  

\begin{figure}\centering
\begin{minipage}[c]{0.32\textwidth}
\includegraphics[width=1.8in, height= 1.5in]{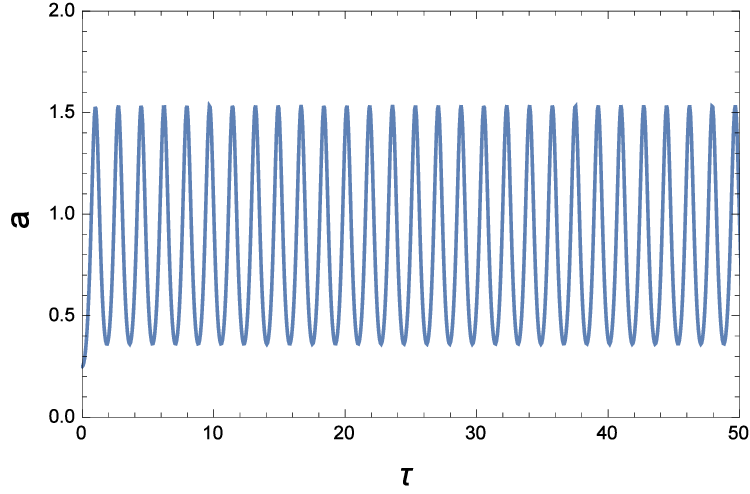}
\end{minipage}%
\begin{minipage}[c]{0.32\textwidth}
\includegraphics[width=1.8in, height= 1.5in]{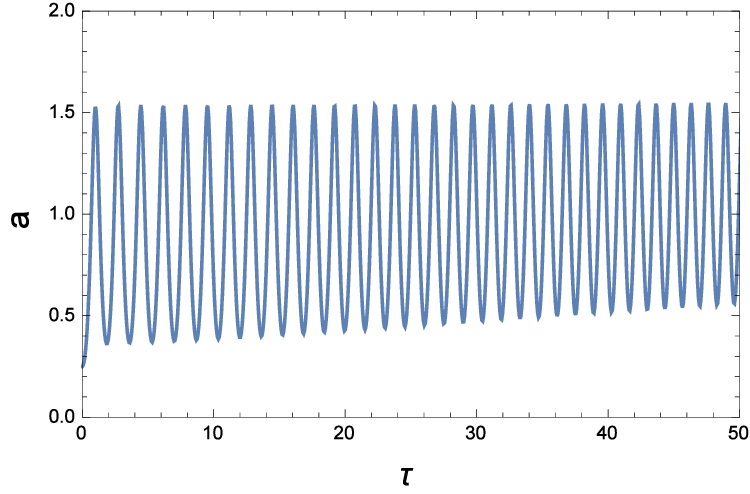}
\end{minipage}%
\begin{minipage}[c]{0.32\textwidth}
\includegraphics[width=1.8in, height= 1.5in]{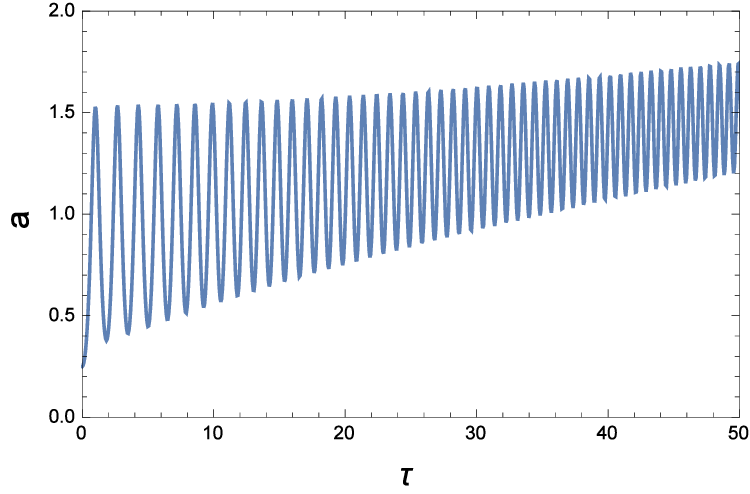}
\end{minipage}
      \caption{(Colour online) The optical-field amplitude $a$ as a function of $\tau$ for $\gamma = 5.5$, $\beta=1.2$, $\alpha=0.5$, $\omega=5$, $\vartheta= - 0.08$, $P= 1.5$, $\phi_0 = -\pi$, $q=1$, $f_0=0.05$, $B = 0.2$, $A = 0.25$. Left graph is for $\lambda=0$, middle graph for $\lambda=-0.01$, and right graph for $\lambda=-0.05$.}{\label{fig3a}}
  \end{figure}

\begin{figure}\centering
\begin{minipage}[c]{0.32\textwidth}
\includegraphics[width=1.8in, height= 1.5in]{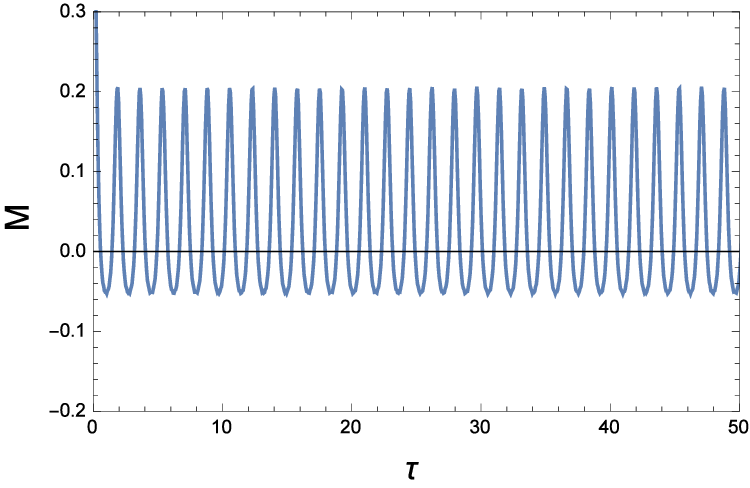}
\end{minipage}%
\begin{minipage}[c]{0.32\textwidth}
\includegraphics[width=1.8in, height= 1.5in]{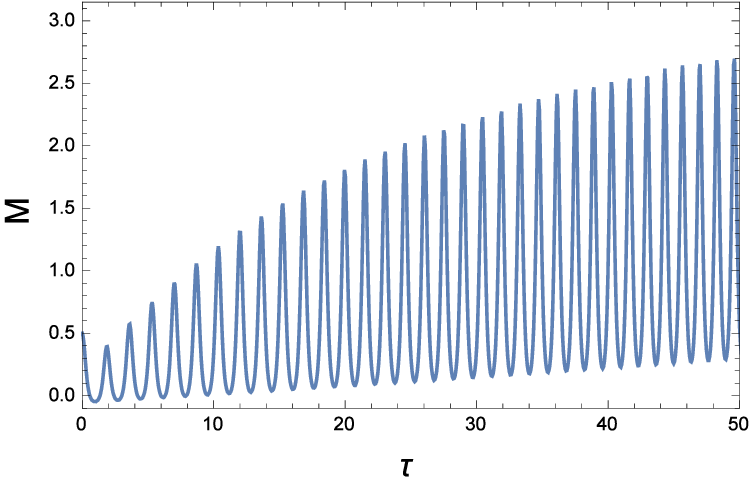}
\end{minipage}%
\begin{minipage}[c]{0.32\textwidth}
\includegraphics[width=1.8in, height= 1.5in]{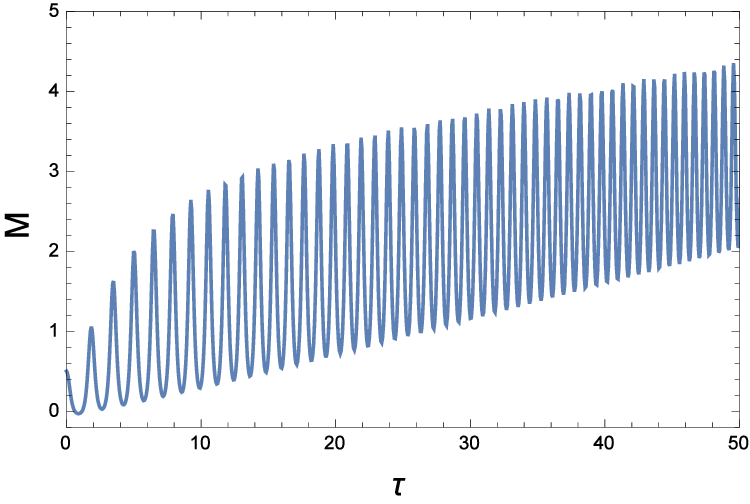}
\end{minipage}
      \caption{(Colour online) Instantaneous frequency $M$ of the optical field, plotted as a function of $\tau$ for $\gamma = 5.5$, $\beta=1.2$, $\alpha=0.5$, $\omega=5$, $\vartheta= - 0.08$, $P= 1.5$, $\phi_0 = -\pi$, $q=1$, $f_0=0.05$, $B = 0.2$,
$A = 0.25$. Left graph is for $\lambda=0$, middle graph for $\lambda=-0.01$, and right graph for $\lambda=-0.05$.}{\label{fig3b}}
  \end{figure}

When $A$ is decreased to $0.25$ with $B=0.2$, figs. \ref{fig3a} and \ref{fig3b} suggest that the optical-field amplitude $a(\tau)$ and instantaneous frequency $M(\tau)$ are qualitatively unaffected. However, it is quite apparent that the amplitude of dissipative soliton crystals is smaller for this value of $A$, implying that an increase of the thermal parameter $A$ favours the increase of the soliton crystal amplitude. Instructively the latter behaviour of the soliton-crystal amplitude is observed if both $A$ and $B$ are fixed but the average power $P$ is decreased, as eq. (\ref{eq9c}) attests.  

\par
\begin{figure}\centering
\begin{minipage}[c]{0.52\textwidth}
\includegraphics[width=2.8in, height= 2.in]{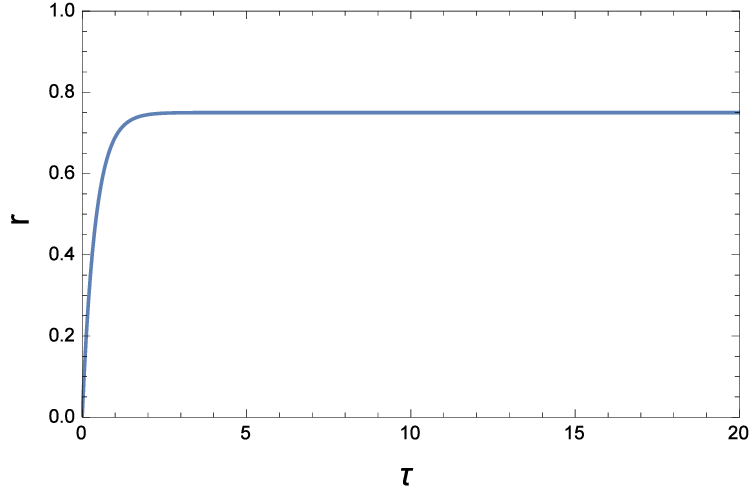}
\end{minipage}%
\begin{minipage}[c]{0.52\textwidth}
\includegraphics[width=2.8in, height= 2.in]{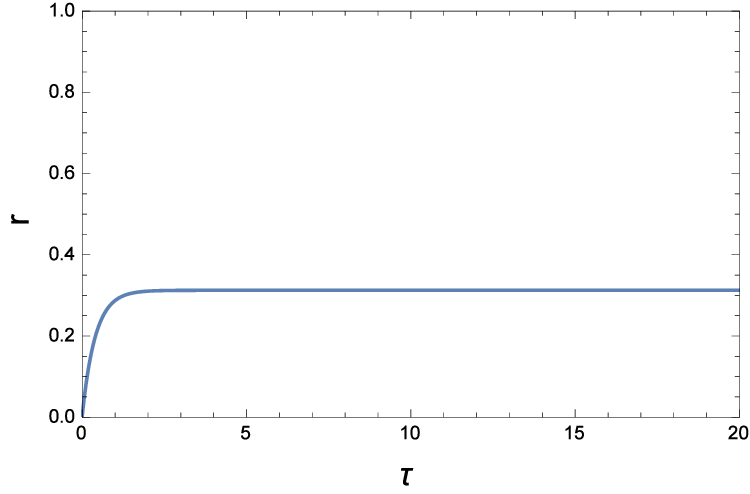}
\end{minipage}
      \caption{(Colour online) The reduced temperature difference $r=(T-T_0)/T_0$ plotted as a function of $\tau$, for $qT_0=5$, $\vartheta=-0.08$, $P=1.5$ and $B=0.2$. Left graph is for $A = 0.5$, right graph is for $A= 0.25$.}{\label{fig4}}
  \end{figure}

 A peculiar feature of the model under study, that emerged from numerical simulations of the coupled set eqs. (\ref{eq9a1})-(\ref{eq9c}), is the observation that the roundtrip motion of the optical-field amplitude does not affect the temperature variation in the microresonator cavity. On this last point, the two graphs in figure \ref{fig4} are profiles of the reduced temperature difference $r(\tau)$ versus $\tau$, for $B=0.2$, $A=0.5$ (left graph) and $B=0.2$, $A=0.25$ (right graph). Changing $\lambda$, or any other parameter in the Lugiato-Lefever equation (\ref{eq1}), does not affect these profiles either qualitatively or quantitatively. 
The two graphs in figure \ref{fig4} suggest an exponential increase of temperature in the microresonator cavity from $T_e$, and its quick saturation to some finite threshold value during the soliton-crystal roundtrips. As we increase $A$ with $B$ fixed, this threshold value is increased. We observed in our numerical simulations that when $A$ is chosen smaller than $B$, the cavity temperature instead decreases during the optical-field roundtrips but also saturates to a finite lower threshold temperature, after some roundtrips.

\section{Conclusion}
In this study we considered a model for soliton-comb propagation in optical Kerr microresonators, in the presence of thermo-optic effects. The model \cite{mod1,mod2} consists of a Lugiato-Lefever equation with an extra detuning referred to as thermal detuning and accounting for thermo-optic effects, coupled to the heat equation. This latter equation is characterized by two parameters namely $A$ and $B$, assumed to control temperature variation in the microresonator cavity during the optical field roundtrips, as well as a source term proportional to the average power of the optical field within the cavity. By solving numerically the model equations assuming a specific ansatz \cite{fand,ans1,ans2,ans3,ans4} for the optical field, we found that an increase of the characteristic parameter $A$ with $B$ fixed favours an increase of the amplitude of dissipative soliton crystals. We noted that the temperature profile was not influenced by characteristic parameters in the optical-field equation, but only by those involved in the heat equation. We observed that the temperature of the microresonator cavity increases exponentially after some roundtrips of the optical field in the cavity, and quickly saturates to a threshold temperature which is increased with an increase of $A$, fixing the other coefficient (i.e. $B$). \par As we indicated the spatio-temporal profile of dissipative soliton crystals does not influence the temperature variation in the microresonator cavity, actually this is the main peculiarity of the model. This peculiarity comes from the fact that in the heat equation (\ref{eq2}), the source term is proportional to the average power $P$ of the optical field. 

\ack
Work partially done at ICTP, Trieste Italy. A. M. Dikand\'e and I. Ndifon Ngek thank the Ministry of Higher Education of
Cameroon (MINESUP), for financial assistance within the
framework of the "Research
Modernization" Allowances.

\section*{ORCID iD}
Alain M Dikandé: https://orcid.org/0000-0002-7577-6853




\section*{References}

\begin{thebibliography}{99}
\bibitem{a1} Udem T, Holzwarth R and H\"ansch T W 2002 {\it Nature} {\bf 416} 233.
\bibitem{a2}Jones D J, Diddams S A, Ranka J K et al. 2000 {\it Science} \textbf{288} 635.
\bibitem{a3}Del’Haye P, Schliesser A, Arcizet O, Wilken T, Holzwarth R and
Kippenberg T J 2007 {\it Nature} \textbf{450} 1214.
\bibitem{a4} Kippenberg T J, Holzwarth R and Diddams S A 2011 {\it Science} \textbf{332} 555.
\bibitem{a5} Chembo Y K 2016 {\it Nanophotonics} \textbf{5}
214.
\bibitem{a6}Lugiato L A, Prati F, Gorodetsky M L and Kippenberg J T
2018 {\it Philos. T. R. Soc.} A \textbf{376} 20180113.
\bibitem{a61}Ricardo E O, Bertoni-Ocampo C, Maldonado-Terr\'on M, Zurita A S, Ramirez-Alarcon R, Ramirez H C, Castro-Beltr\'an R and U'Ren A B 2021 {\it Photonics Research} \textbf{9} 2237.
\bibitem{a62}Kovach A, Chen D, He J,
Choi H, Dogan A H, Ghasemkhani M, Taheri H and Armani A M 2020 {\it Advances in Optics and Photonics} \textbf{12} 135.
\bibitem{a7}Qing L, Briles T C, Westly D A,
Drake T E, Stone J R, Ilic B R, Diddams S A, Papp S B and Srinivasan K 2017 {\it Optica} \textbf{4} 193.
\bibitem{a8}Pfeiffer M H P, Herkommer C, Liu J, Guo H, Karpov M, Lucas E, Zervas M and Kippenberg T J 2017 {\it Optica} \textbf{4} 684.
\bibitem{a9}Cole D C, Lamb E S, Del’Haye P, Diddams S A and Papp S B 2017 {\it Nature
Photonics} \textbf{11} 671.
\bibitem{a10} Qi Z, D’Aguanno G and Menyuk C R 2017  {\it J. Opt. Soc. Am.} B \textbf{34} 785.
\bibitem{a11} Wang W, Lu Z, Zhang W, Chu S T, Little B E, Wang L, Xie X, Liu M, Yang Q, Wang L, Zhao J, Wang G, Sun Q, Liu Y, Wang Y and Zhao W 2018 {\it Opt. Lett.}
\textbf{43} 2002.
\bibitem{a12}Dikand\'e Bitha R D and Dikand\'e A M 2019 {\it Eur. Phys. Jour.} D\textbf{73} 152.
\bibitem{a13}Dikand\'e Bitha R D and Dikand\'e A M 2018 {\it Phys. Rev.} A\textbf{97} 033813.
\bibitem{d1}Dikand\'e A M 2010 {\it Phys. Rev.} A \textbf{81} 013821.
\bibitem{d2}Jubgang Fandio D Jr, Dikand\'e A M and Sunda Meya A 2015 {\it Phys. Rev.} A \textbf{92} 053850.
\bibitem{fw1}Herr T, Hartinger K, Riemensberger J, Wang C Y, Gavartin E, Holzwarth E, Gorodetsky M L and Kippenberg J T 2012 {\it Nature Photonics} \textbf{6} 480.
\bibitem{lid}Trocha P et al.2018 {\it Science} \textbf{359} 887.
\bibitem{lid1}Obrzud E et al. 2019 {\it Nature Photonics} \textbf{13} 31.
\bibitem{lid2}Spencer D T et al. 2018 {\it Nature} \textbf{557} 81.
\bibitem{10}Godey C, Balakireva I V, Coillet A and Chembo Y K 2014
{\it Phys. Rev.} A \textbf{89} 063814.
\bibitem{11}Carmon T, Yang L and Vahala K J 2004 {\it Optics Express} \textbf{12} 4742.
\bibitem{12}Herr T, Brasch V, Jost J D, Wang C Y, Kondratiev
N M, Gorodetsky M L, Kippenberg T J 2014 {\it Nature Photonics} \textbf{8} 145.
\bibitem{mod1}Qi Z , Leshem A, Jaramillo-Villegas J, D’Aguanno G, Carruthers T F, Omri Gat, Weiner A M and Menyuk C R 2020 {\it Optics Express} \textbf{28} 36304.
\bibitem{mod2}Lobanov V E, Kondratiev N M and Bilenko I A 2021 {\it Opt. Lett.} \textbf{46} 2380.
\bibitem{fand}Jubgang Fandio D Jr, Dikand\'e A M and Sunda Meya A 2021 {\it J. Opt. Soc. Am.} B \textbf{37} A175.
\bibitem{ans1}Akeweje E O, Bader G, Dikand\'e A M and Kameni Nteutse P 2021 {\it J. Mod. Opt.} \textbf{68} 1211.
\bibitem{ans2}Kameni Nteutse P, Dikand\'e A M and Zekeng S 2021 {\it J. Opt.} \textbf{23} 035402.
\bibitem{ans3}Ntongwe Mesumbe E and Dikand\'e A M 2019
{\it Opt. Quantum Elec.} \textbf{51} 361.
\bibitem{ans4}Achankeng Leke P and Dikand\'e A M 2020 {\it Appl. Phys.} B \textbf{126} 157.

\end{thebibliography}
\end{document}